\documentclass[aps,preprintnumbers,nofootinbib]{revtex4}
\usepackage{graphicx}
\usepackage{amssymb}
\usepackage{amsmath,mathrsfs,verbatim}
\usepackage{times}
\usepackage{latexsym}
\def\beq{\begin{equation}}
\def\eeq{\end{equation}}
\def\bey{\begin{eqnarray}}
\def\eey{\end{eqnarray}}

\def\lsim{\mathrel{\raise.3ex\hbox{$<$\kern-.75em\lower1ex\hbox{$\sim$}}}}
\def\gsim{\mathrel{\raise.3ex\hbox{$>$\kern-.75em\lower1ex\hbox{$\sim$}}}}

\newcommand{\be}{\begin{equation}}
\newcommand{\ee}{\end{equation}}

\newcommand{\gev}{{\rm ~GeV }}

\newcommand{\diracslash}[1]{#1\;\!\!\!\!\!/}
\newcommand{\cms}{{\rm cm}^3/{\rm s}}

\begin{document}

\preprint{MCTP-12-18}

\title{Three Exceptions for Thermal Dark Matter with Enhanced Annihilation to $\gamma \gamma$}

\author{Sean Tulin, Hai-Bo Yu, and Kathryn M. Zurek}
\affiliation{Department of Physics, University of Michigan, Ann Arbor, MI 48109  }

\date{\today}

\begin{abstract}

Recently, there have been hints for dark matter (DM) annihilation in the galactic center to one or more photon lines.  
In order to achieve the observed photon line flux, DM must have a relatively large effective coupling to photons, typically generated radiatively from large couplings to charged particles.  When kinematically accessible, direct annihilation of DM to these charged particles is far too large to accommodate both the DM relic density and constraints from the observed flux of continuum photons from the galactic center, halo and dwarf galaxies.  We discuss three exceptions to these obstacles, generating the observed line signal while providing the correct relic density and evading photon continuum constraints.  The exceptions are (i) coannihilation, where the DM density is set by interactions with a heavier state that is not populated today, (ii) forbidden channels, where DM annihilates to heavier states that are kinematically blocked today, but open in the early Universe, and (iii) asymmetric DM, where the relic density is set by a primordial asymmetry.  We build simple models to realize these scenarios. 

\end{abstract}

\maketitle

\section{Introduction} 

Dark Matter (DM) is one of the primary pieces of evidence for physics beyond the Standard Model (SM).   Although its particle physics nature remains a mystery, in many theories DM is a weakly interacting massive particle (WIMP), whose its relic density is determined by annihilation to SM particles with a weak-scale cross section.  The same annihilation processes that sets its density can give rise to observable photon signals that can be observed in the Universe today.  Radiation from DM annihilation into electrically charged particles produces an additional component to the continuous $\gamma$ spectrum.  More strikingly, DM can annihilate directly into $\gamma \gamma$, $\gamma Z$, or $\gamma h$ through processes with charged particles in loops, producing monoenergetic $\gamma$ lines.  Since $\gamma$ lines are not easily mimicked by astrophysical backgrounds, they are a ``smoking gun'' signature for DM.

Recently, several groups~\cite{Weniger:2012tx,Bringmann:2012vr,Tempel:2012ey,Su:2012ft} have reported a $\gamma$ line spectral feature at $E_\gamma \approx 130$ GeV in publicly available data from the Fermi Large Area Telescope (LAT)~\cite{Atwood:2009ez}.  It is an exciting possibility that WIMP DM may explain this signal.  The required annihilation cross section to $\gamma \gamma$ is $\langle \sigma v\rangle_{\gamma \gamma} \approx 10^{-27} \: \cms$ depending on the DM profile, an order of magnitude smaller than needed for the relic density~\cite{Weniger:2012tx,Tempel:2012ey}. Moreover, Ref.~\cite{Su:2012ft,Su:2012zg} reported evidence for an additional $\gamma$ line at $E_\gamma \approx 111$ GeV; the pair of lines is kinematically consistent with DM with mass $m_\chi \approx 130 \; {\rm GeV}$ annihilating to both $\gamma \gamma$ and $\gamma Z$, or $m_\chi \approx 140 \; {\rm GeV}$ annihilating to both $\gamma Z$ and $\gamma h$, as in the model of~\cite{Bertone:2009cb,Jackson:2009kg,Bertone:2010fn}.  The Fermi collaboration has not confirmed these results, and null results from their most recent $\gamma$ line search~\cite{Ackermann:2012qk} are in tension with the cross section required to explain this signal.\footnote{Note the Fermi collaboration searched for lines using all sky gamma-ray maps~\cite{Ackermann:2012qk}, and the results are obtained for $|b|>10^{\rm o}$ plus a $20^o\times20^{\rm o}$ square at the galactic center, using Pass 6 processing. However, in the analysis by Ref.~\cite{Weniger:2012tx}, the search regions were optimized for DM signals and Pass 7 was used. }  
Also, the $\gamma\gamma$ cross section is consistent with null searches for $\gamma$ lines from dwarf galaxies~\cite{GeringerSameth:2012sr}.  Astrophysical sources have been suggested to explain the signal~\cite{Profumo:2012tr, Aharonian:2012cs,Boyarsky:2012ca}, though Refs.~\cite{Su:2012ft,Hektor:2012kc, Yang:2012ha} suggest that the signal may prefer a DM interpretation if it is not due to instrumental effects.  In any case, further analysis with more data is required before claiming a definitive discovery of DM.

\begin{figure}[t]
\includegraphics[scale=1]{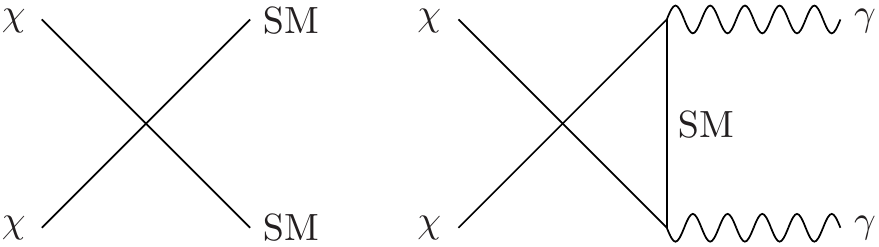}
 \caption{WIMP annihilation to charged SM final states ({\it Left}), {\it e.g.}, fermions $f \bar f$ or $WW$, generates annihilation to $\gamma\gamma$ at one-loop ({\it Right}).}
 \label{fig:ggff}
 \end{figure}

Since the electric charge of DM is zero (or extremely tiny~\cite{McDermott:2010pa}), $\gamma$ couplings to DM arise radiatively.  For example, in many WIMP models, DM couples to SM charged particles ({\it e.g.}, fermion pairs $f \bar f$ or $WW$) through weak-scale mediators, giving a tree-level annihilation channel $\chi \chi \to f \bar f$ or $WW$. In this case, $\chi \chi \to \gamma \gamma$ arises at one-loop, through virtual charged SM particles as shown in Fig.~\ref{fig:ggff}.  From the effective field theory point of view, these one-loop processes lead to a dimension six operator $|\chi|^2F_{\mu\nu}F^{\mu\nu}$ for scalar DM and dimension seven operator $\bar{\chi}\gamma_5\chi F_{\mu\nu}F^{\mu\nu}$ for fermionic DM, where $F_{\mu\nu}$ is the electromagnetic field strength~\cite{Goodman:2010qn,Rajaraman:2012db, Abazajian:2011tk}.
Therefore, one expects the enhancement of DM annihilation to charged SM states over $\gamma\gamma$ to scale as 
\beq \label{naive}
\langle \sigma v \rangle_{f \bar f, WW}/\langle \sigma v \rangle_{\gamma \gamma} \sim (\pi/\alpha)^2 \approx 10^5.
\eeq 
Some enhancement of the photon signal can be achieved by placing the charged virtual particles and DM in $SU(2)_L$ multiplets (see, {\it e.g.}, Ref.~\cite{Acharya:2012dz}), though the ratio remains large.  

Due to the large ratio in Eq.~\eqref{naive}, a DM explanation for the $\gamma$ line signal faces two main obstacles.  First, Fermi LAT observations place strong constraints on DM annihilation to $f \bar f$ or $WW$ from the $\gamma$ continuum, at the level of $\langle \sigma v \rangle_{f \bar f, WW} \lesssim \mathcal{O}({\rm few}) \times 10^{-25} \: \cms$, depending on the final state particles~\cite{GeringerSameth:2011iw,Ackermann:2011wa, Ackermann:2012qk}.  
As a result, one na\"{i}vely expects an upper bound on the $\gamma\gamma$ annihilation cross-section $\langle \sigma v \rangle_{\gamma\gamma} \lesssim \mathcal{O}({\rm few}) \times 10^{-30} \; \cms$, which is well below what is needed to generate the Fermi signal. Indeed, the neutralino interpretation of the line signal has been disfavored by these arguments~\cite{Buchmuller:2012rc,Cohen:2012me,Cholis:2012fb}.

Second, the total annihilation cross section in the early Universe must be $\langle \sigma v \rangle \sim 3 \times 10^{-26} \: \cms$ to generate the observed relic density.  For $\langle \sigma v \rangle_{\gamma\gamma} \approx 10^{-27} \: \cms$, according to Eq.~\eqref{naive}, one expects $\chi \chi \to f\bar f$ or $WW$ to be far too large, giving a relic density much smaller than observed.  Even if tree-level annihilation is $p$-wave suppressed, the additional $\mathcal{O}(10)$ suppression from the DM relative velocity (squared) is not sufficient to avoid depleting the DM relic density.

In addition to the dimension six or seven operators just discussed, fermionic DM may couple to photons through a dimension-five magnetic dipole operator ${\bar{\chi}}\sigma^{\mu\nu}\chi F_{\mu\nu}$ or electric dipole operator ${\bar\chi}\sigma^{\mu\nu}\gamma_5\chi F_{\mu\nu}$, where $\sigma^{\mu\nu}\equiv-i [\gamma^\mu,\gamma^\nu]/2$.  
This type of DM can be found in models where DM is a composite state~\cite{Bagnasco:1993st,Pospelov:2000bq,Sigurdson:2004zp,Gardner:2008yn,Masso:2009mu,Cho:2010br,An:2010kc,Chang:2010en,Banks:2010eh},
and was considered recently in connection with the Fermi line signal~\cite{Weiner:2012cb}.  Dipolar DM encounters similar challenges in explaining both the line signal and relic density, since the dipole operator mediates $\chi \bar \chi \to f \bar f$ as well as $\chi \bar \chi \to \gamma \gamma$.  For the magnetic dipole case, fixing $\langle \sigma v \rangle_{\gamma \gamma} = 10^{-27} \: \cms$ gives $\langle \sigma v \rangle_{f \bar f} \gtrsim 10^{-25} \: \cms$, which gives a too-small DM relic density. In the electric dipole case, $\chi \bar \chi \to f \bar f$ is $p$-wave suppressed, and the relic density is too large, unless there are additional annihilation channels.  Furthermore, Dirac DM models with such large dipole interactions are excluded by direct detection experiments~\cite{Fortin:2011hv}.

So far, we have seen both the relic density constraint and the continuum photon bound strongly disfavor simple WIMP models  for enhanced $\gamma$ line signals.  To alleviate these tensions, we have to consider extensions to the simple WIMP models with designed features to enhance the $\gamma\gamma$ signal~\cite{Cline:2012nw,Choi:2012ap,Lee:2012bq, Buckley:2012ws, Chu:2012qy,Das:2012ys,Kang:2012bq,Acharya:2012dz,Feng:2012gs,Weiner:2012cb,Heo:2012dk,Frandsen:2012db,Kyae:2012vi,Park:2012xq,Dudas:2012pb,Dudas:2009uq,Mambrini:2009ad}.

In this paper, we discuss three generic scenarios that are exceptions to these constraints, allowing for a large $\gamma \gamma$ annihilation rate while annihilation to fermions is suppressed compared to Eq.~\eqref{naive}, both in the early Universe and in the galactic halo today.
The three exceptions are:
\begin{itemize}
\item {\it Coannihilation:} The relic density is set by $\chi_1 \chi_2 \rightarrow f \bar f$, where $\chi_1$ is DM and $\chi_2$ is a next-to-lightest state nearby in mass.  Annihilation to $f \bar f$ is suppressed during freeze-out by the $\chi_1$-$\chi_2$ mass gap, giving the correct relic density for $\mathcal{O}(10 \: {\rm GeV})$ splitting.  No annihilation to $f \bar f$ occurs today since $\chi_2$ decays to $\chi_1$ and is not populated.
\item {\it Forbidden channels:} DM annihilates to charged fermions $F \bar F$ that are slightly heavier than the DM particles themselves.  Due to the high velocity tail of the DM distribution, annihilation occurs in the early Universe, setting the relic density, but is kinematically forbidden today.
\item {\it Asymmetric DM (ADM):} The relic density is set by a primordial DM asymmetry, where a large annihilation rate $\chi \chi^\dagger \to f \bar f$ is quenched by the DM chemical potential.  After freeze-out, the asymmetry is washed out by DM particle-antiparticle oscillations due to tiny DM number-violating mass terms.  $\chi \chi^\dagger \to \gamma \gamma$ annihilation can occur today with a large rate, while $\chi \chi^\dagger \to f \bar f$ can be $p$-wave or chirality-suppressed.
\end{itemize}
In the remainder of this work, we study in detail several minimal DM models as examples to illustrate each of these mechanisms.  In each case, we show that an enhanced $\gamma\gamma$ annihilation rate can naturally be reconciled with the observed relic density and present $\gamma$ continuum constraints.

In Sec.~\ref{sec:coann}, we discuss coannihilation, presenting two models: (i) magnetic dipolar DM, and (ii) coannihilation with charged partners, which generates DM coupling to $\gamma\gamma$ at dimension seven.  In Sec.~\ref{sec:forbidden}, we consider a model with forbidden channels, and we derive the mass gap between DM particles and charged states required for the correct thermal relic density.  In Sec.~\ref{sec:adm}, we present a scalar ADM model and discuss the ingredients necessary for generating the $\gamma$ line while remaining consistent with other constraints.   Our conclusions are summarized in Sec.~\ref{sec:conclusions}.  We focus in this paper on models needed to explain the 130 GeV line, though we emphasize that our results are easily generalized to the case of multiple lines.

\section{Coannihilation}
\label{sec:coann}

In coannihilation scenarios, DM freeze-out is dominated by annihilation with a next-to-lightest state that is nearby in mass.  For concreteness, we consider $\chi_1 \chi_2 \to f \bar{f}$, where $\chi_1$ is the DM, $\chi_2$ is the nearby state, and $f$ is a SM fermion.  We assume that the $\chi_1 \chi_2$ coannihilation channel is dominant in the early Universe, while direct $\chi_1 \chi_1$ annihilation is suppressed.  If the mass splitting $\Delta m \equiv m_{2} - m_{1}$ is comparable to the freeze-out temperature $T_f$, coannihilation can provide a natural framework for enhanced $\gamma$ signals from thermal DM:
\begin{itemize}
\item In the early Universe, the thermally-averaged coannihilation cross section is suppressed by a Boltzmann factor $\exp(-\Delta m/T)$.  For $\Delta m \sim T_f$, the coannihilation rate becomes moderately suppressed, requiring larger couplings to reproduce the correct thermal relic density.  
\item In the present Universe, $\chi_2$ is not populated, and therefore $\chi_1 \chi_2 \to f \bar{f}$ does not contribute to any indirect detection signals.  However, direct annihilation $\chi_1 \chi_1 \to \gamma \gamma$ can occur, and the rate can be enhanced due to the large couplings required for thermal freeze-out.
\end{itemize}
Ultimately, within a given model, there will exist a preferred parameter region for $\Delta m$ and couplings that can simultaneously explain the relic DM density and the observed $\gamma$ signal.  In this section, we first discuss some preliminaries for computing the DM relic density, closely following Ref.~\cite{Griest:1990kh}, and then we consider specific models in parts A and B.

Similar to single species freeze-out, the relic DM abundance for a general coannihilation scenario is computed by solving a Boltzmann equation
\be 
\dot{n}_\chi + 3 H n_\chi = - \langle \sigma_{\rm eff} v \rangle \big(n_\chi^2 - (n_\chi^{\rm eq})^2\big) \label{eq:Boltzmann}
\ee
where $n_\chi \equiv \sum_i n_{\chi_i}$ is the total $\chi_{i}$ density.  In writing Eq.~\eqref{eq:Boltzmann} in terms of only $n_\chi$, we assume the individual densities $n_{\chi_{i}}$ are in chemical equilibrium due to rapid $\chi_i f \leftrightarrow \chi_j f$ and $\chi_i \leftrightarrow \chi_j f \bar{f}$ processes, such that
\be \label{rdef}
\frac{n_{\chi_i}}{n_{\chi}} \approx \frac{n_{\chi_i}^{\rm eq}}{n_{\chi}^{\rm eq}} 
=\frac{g_i (1+\Delta_i)^{3/2}\exp(-x\Delta_i)}{g_{\rm eff}}  \equiv r_i \, .
\ee
We have defined $x \equiv m_1/T$, $\Delta_i \equiv (m_{i} - m_{1})/m_{1}$, and $g_{\rm eff}\equiv \sum_{i} g_i (1+\Delta_i)^{3/2}\exp(-x\Delta_i)$, with $g_{i}$ degrees of freedom for $\chi_i$.  The thermally-averaged effective cross section is $\langle \sigma_{\rm eff}v \rangle \equiv \sum_{i,j} r_i r_j \langle \sigma_{ij} v \rangle$, where $\sigma_{ij}$ is $\chi_i \chi_j$ annihilation cross section and its thermal average is
\be
\langle \sigma_{ij} v \rangle = \frac{x^{3/2}}{2\sqrt{\pi}} \int_0^\infty dv \, v^2 \, (\sigma_{ij}v) \, e^{- v^2 x/4} \; . \label{velavg}
\ee
The DM relic density today is given by
\be
\Omega_{\rm dm} h^2=\frac{1.07\times10^9 \, {\rm GeV}^{-1}}{g^{1/2}_* m_{\rm Pl}\left[\int^\infty_{x_f}x^{-2}\left<\sigma _{\rm eff} v\right> \, dx\right]} \; , \label{relic}
\ee
where $m_{\rm Pl}\approx 1.22\times10^{19}~{\rm GeV}$ is the Planck mass and $g_*$ is the number of degrees of freedom in the thermal bath during freeze-out.
The freeze-out temperature $T_f = m_1/x_f$ is obtained by solving $x_f = \ln \big( 0.038 \, g_{\rm eff} m_1 m_{ \rm Pl} \left<\sigma_{\rm eff} v\right>/ \sqrt{g_* x_f} \big)$, which can be done iteratively.  Alternately, one can directly solve Eq.~\eqref{eq:Boltzmann} numerically; for the cases we consider below, we find that the agreement with Eq.~\eqref{relic} is better than $\sim1-3\%$ depending on the mass splitting.

Now, we discuss two models which give rise to the Fermi line signal and a correct relic density with the coannihilation effect in the early Universe.\footnote{To be clear, our models rely on the mass splitting $\Delta m$ to suppress $\langle \sigma_{\rm eff} v\rangle$, which is dominated by large $\chi_1 \chi_2$ and $\chi_2 \chi_2$ annihilation cross sections.  This is distinct from models where $\chi_1\chi_1$ annihilation is itself too large, and $\langle \sigma_{\rm eff} v\rangle$ can be suppressed by $1/g_{\rm eff}$ by having a ``parasitic'' species $\chi_2$ that does not annihilate strongly (see, {\it e.g.},~\cite{Profumo:2006bx,Feldman:2009wv}).} 

\subsection{Magnetic dipolar dark matter}

\begin{figure}[t]
\includegraphics[scale=1]{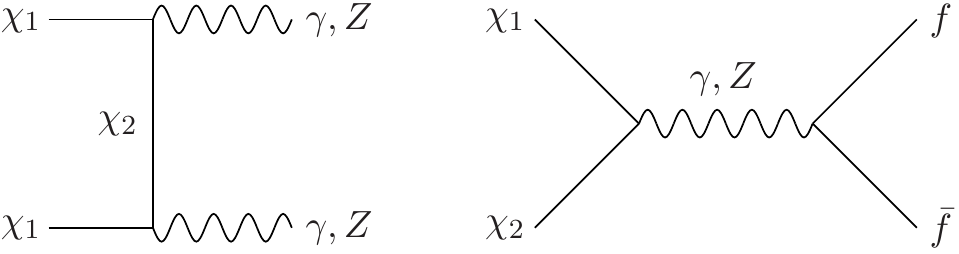}
\caption{Magnetic dipolar DM $\chi_1$ annihilates to $\gamma\gamma, \gamma Z, ZZ$ ({\it Left}), while $f \bar f$ occurs by coannihilation only with $\chi_2$ ({\it Right}).}
\label{feyn2}
\end{figure}

Although the electric charge of DM must be zero or very small, DM can possess a sizable electromagnetic interaction through an electric or magnetic dipole moment~\cite{Bagnasco:1993st,Pospelov:2000bq,Sigurdson:2004zp}.  As we show, magnetic dipolar DM can account for the Fermi $\gamma$ signal, and coannihilation plays an essential role in achieving the correct DM relic density.\footnote{The case of purely {\it electric} dipolar DM cannot explain the Fermi $\gamma$ line, since the coannihilation process $\chi_1 \chi_2 \to f \bar f$ setting the DM relic density is $p$-wave suppressed~\cite{Sigurdson:2004zp}.  Fixing the electric dipole moment to require $\sigma(\chi_1 \chi_1 \to \gamma \gamma)v \approx 10^{-27}$ cm$^3/$s, the DM relic density is too large (even if $\Delta m =0$) unless additional annihilation channels are present.} We consider a Dirac fermion $\chi$ coupled to the hypercharge field strength $B_{\mu\nu}$ through a magnetic dipole interaction, with Lagrangian 
\be
\mathscr{L}= i \bar\chi \diracslash{\partial}  \chi +m_D \bar{\chi}\chi+\frac{m_M}{2}\left(\bar \chi^c {\chi}+\bar{\chi} \chi^c \right) + \frac{\mu_B}{2}\bar{\chi}\sigma^{\mu\nu}\chi B_{\mu\nu} \, ,
\label{eq:lag}
\ee
where $\chi^c=-i\gamma^2\chi^*$ is the charge-conjugated $\chi$ field.  We have two mass terms: a Dirac mass $m_D$, and a Majorana mass $m_M$, which splits $\chi$ into two Majorana fermions $\chi_{1,2}$ with masses $m_{1,2} = |m_D \pm m_M|$. Taking $m_1<m_2$, $\chi_1$ is the DM.  In terms of $\chi_{1,2}$, the magnetic dipole interaction becomes
\be
\mathscr{L}_{\rm int} = \frac{\mu_\gamma}{2} \bar{\chi}_2 \sigma^{\mu\nu} \chi_1 F_{\mu\nu} + \frac{\mu_Z}{2} \bar{\chi}_2 \sigma^{\mu\nu} \chi_1 Z_{\mu\nu},
\ee
where $\mu_\gamma = \mu_B c_W$ and $\mu_Z=-\mu_B s_W$, and $s_W$ ($c_W$) is the (co)sine of the weak mixing angle.  For Majorana states, only $\chi_1 \leftrightarrow \chi_2$ transition dipole moments are allowed.  The photon and $Z$ boson field strengths are $F_{\mu\nu}$ and $Z_{\mu\nu}$, respectively.

DM can annihilate to $\gamma\gamma$, $\gamma Z$, and $ZZ$ final states, through $t$-channel $\chi_2$ exchange, shown in Fig.~\ref{feyn2}.  The cross sections are
\begin{subequations} \label{eq:ann}
\begin{align}
\sigma (\chi_1 \chi_1 \to \gamma \gamma)v &= \frac{\mu_\gamma^4 m_1^4 m_2^2}{\pi (m_1^2+m_2^2)^2} ,\\
\sigma (\chi_1 \chi_1 \to \gamma Z)v &= \frac{\mu_\gamma^2 \mu_Z^2 (4 m_1^2 - m_Z^2)^3(4 m_1 m_2 + m_Z^2)^2}{128 \pi m_1^4 (2m_1^2+2 m_2^2 - m_Z^2)^2}\\
\sigma (\chi_1 \chi_1 \to ZZ)v &= \frac{\mu_Z^4 (m_1^2 - m_Z^2)^{3/2} (2 m_1 m_2+m_Z^2)^2}{4 \pi m_1 (m_1^2 + m_2^2 - m_Z^2)^2},
\end{align}
\end{subequations}
where $m_Z$ is the $Z$ boson mass.  To explain the Fermi signal, we fix $m_1 = 130$ GeV and $\langle\sigma v\rangle_{\chi_1 \chi_1 \to \gamma \gamma} \approx 10^{-27}$ cm$^3/$s.  For $m_2 \approx m_1$, the $\gamma Z$ and $ZZ$ cross sections are comparable; in particular, $\chi_1 \chi_1 \to \gamma Z$ generates a second $\gamma$ line at an energy $E_\gamma = m_{1} - m_Z^2/(4m_1) \approx 114$ GeV, which may be indicated in the data~\cite{Su:2012ft}. We estimate the size of the $\mu_B$ required to the line signal
\be
\sigma (\chi_1 \chi_1 \to \gamma \gamma)v \approx10^{-27}~{\rm cm^3/s}\left(\frac{\mu_B}{3.6\times10^{-3}\mu_N}\right)^4\left(\frac{m_1}{130~\gev}\right)^2,
\ee
where $\mu_N \approx 0.161~{\gev^{-1}}$ is the nuclear magneton.

\begin{figure}
\includegraphics[scale=1]{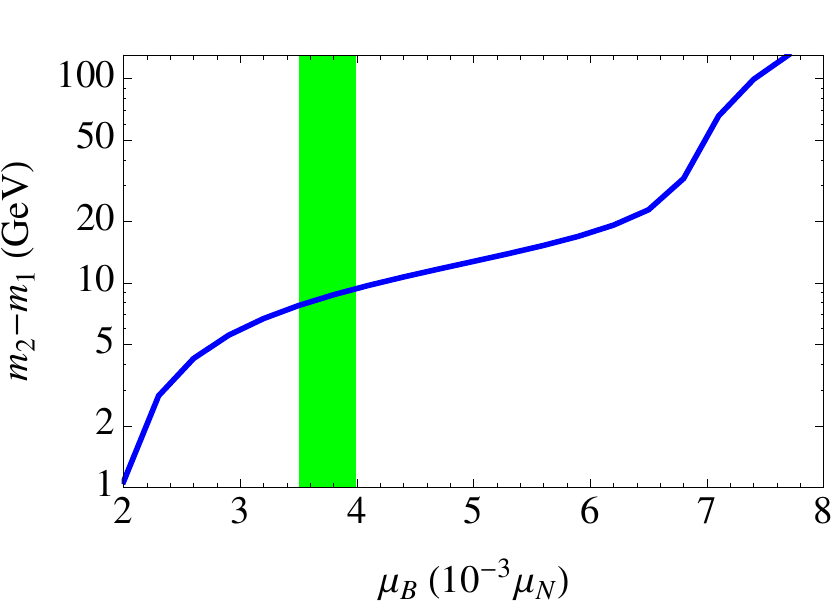}
\includegraphics[scale=1.02]{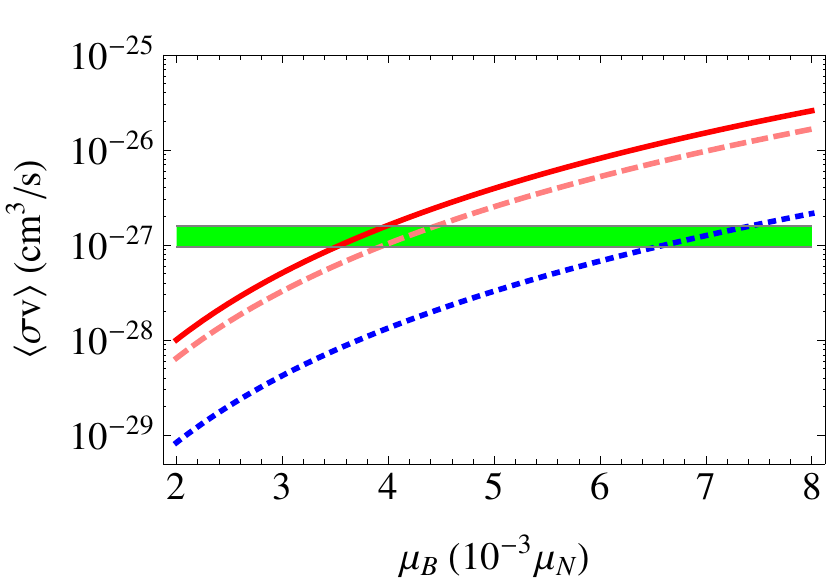}
 \caption{The mass splitting $\Delta m=m_2-m_1$ required for $\Omega_{\chi_1}h^2=0.11$ ({\it Left}) and annihilation cross sections for $\gamma\gamma$ (solid),$\gamma Z$ (dashed) and $ZZ$ (dotted) ({\it Right}) with respect to the dark matter dipole magnitude $\mu_B$. We take $m_1=130~{\rm GeV}$ and the nuclear magnetic magneton $\mu_N\approx0.161~{\rm GeV^{-1}}$. The vertical (horizontal) band on the left (right) panel indicates $\left<\sigma v\right>_{\gamma\gamma}=(1.27\pm0.32)\times10^{-27}~{\rm cm^3/s}$~\cite{Weniger:2012tx}.}
 \label{fig:dipole}
 \end{figure}

In the early Universe, coannihilation $\chi_1\chi_2\rightarrow f\bar f$ provides the dominant annihilation channel, shown in Fig.~\ref{feyn2}.  The cross section is
\be
\sigma (\chi_1 \chi_2 \to f \bar f)v = \alpha c_f \left( \mu_\gamma^2 Q_f^2 + \frac{\mu_\gamma \mu_Z s Q_f (T_{3f} - 2 Q_f s_W^2)}{c_W s_W (s-m_Z^2)}+\frac{\mu_Z^2 s^2 (2 Q_f^2 s_W^4 - 2 Q_f s_W^2 T_{3f} + T_{3f}^2)}{2 c_W^2 s_W^2 (s-m_Z^2)^2} \right)
\label{eq:dpffbar}
\ee
where $Q_f$ is the electric charge in units of $|e|$, $T_{3f}$ is the weak isospin, and $c_f$ is a color factor for fermion $f$ (3 for quarks, 1 for leptons).  The $\chi_1 \chi_2 \to W^+ W^-$ cross section is $\mathcal{O}(1\%)$ of the total $f\bar{f}$ cross section, and can be neglected.  In addition, subleading $\chi_2 \chi_2 \to \gamma \gamma, Z \gamma, ZZ$ also impact the relic density, and the cross sections are obtained by switching $m_1$ and $m_2$ in Eqs.~\eqref{eq:ann}. 

Taking Eq.~(\ref{eq:dpffbar}), we estimate the annihilation cross section $\sigma(\chi_1\chi_2\rightarrow f\bar{f})v\approx1.7\times10^{-25}~{\rm cm^3/s}$ for $m_1=130~\gev$ and $\mu_B\approx3.6\times10^{-3}\mu_N$ as preferred by the Fermi line signal. Clearly, a dipole which is large enough to generate the observed $\gamma \gamma$ line will give rise to too large an annihilation to $f \bar f$ both for the relic density and for continuum constraints in the halo if DM is a Dirac fermion (corresponding to $m_M=0$). This problem is easily solved in a model where the components of the Dirac fermion are split.  In this case, annihilation to fermions proceeds only via $\chi_1 \chi_2 \rightarrow f \bar f$, and the annihilation rate will be suppressed by a Boltzmann factor $\exp(-\Delta m/T_f)$ with $\Delta m=m_2-m_1$. Since $T_f\approx6~\gev$ for $m_1=130~\gev$, we expect $\Delta m\sim {\cal O}(10)~\gev$ for the suppression mechanism to work.\footnote{In contrast, Ref.~\cite{Weiner:2012cb} focused on dipolar DM with $\Delta m \sim \mathcal{O}(100 \: {\rm keV})$, which is sufficient to avoid continuum and direct detection constraints.  Although $\Delta m$ is too small to obtain the correct relic density by coannihilation, they argue that the Fermi line might be reconciled with the DM abundance by having {\it both} electric and magnetic dipole moments, or through momentum-dependent dipole form factors.}

We calculate the relic density of $\chi_1$ numerically by using Eq.~(\ref{relic}). In Fig.~\ref{fig:dipole} ({\it Left}), we show the mass splitting between $\chi_1$ and $\chi_2$ required for the correct DM relic density as a function of the DM magnetic dipole moment $\mu_B$ (solid). We can see that the relic density constraint requires a larger mass splitting for a larger $\mu_B$ as expected. For $\mu_B$ preferred by the Fermi line signal as indicated by the vertical green band, the required mass splitting is $\sim 7-10~\gev$. Note for a large $\mu_B$, the annihilation cross section to $\gamma\gamma$ becomes large enough to set the relic density without the presence of $\chi_2$ in the thermal bath as indicated by the steep rise of the curve for $\mu_B\gtrsim7.5\times10^{-3}\mu_B$ . In Fig.~\ref{fig:dipole} ({\it Right}), we plot annihilation cross sections for $\gamma\gamma$ (solid), $\gamma Z$ (dashed) and $ZZ$ (dotted) with respect to $\mu_B$ for $m_1=130~\gev$. 

In this model, $\chi_2$ decays to $\chi_1$ promptly in the early Universe and it is not populated now due to the mass splitting. Thus, the model evades the continuum photon constraint. Since the preferred $\Delta m$ is too large for signals in direct detection experiments, the most promising way to explore this model is through the Large Hadron Collider (LHC)~\cite{Fortin:2011hv, Weiner:2012cb, Barger:2012pf, Bai:2011jg}.

\subsection{Coannihilation with charged partners}

Next, we present another coannihilation scenario in which the coannihilating state $\chi_2$ carries electric charge.  To be concrete, we consider the following Lagrangian:
\be
\mathscr{L}_{\rm int} = \bar{\chi}_2 ( g_S + g_P \gamma_5) \chi_1 \phi + \bar{f} ( g^\prime_S + g^\prime_P \gamma_5) f^\prime \phi + \textrm{h.c.}
\ee
where $f,f^\prime$ are SM fermions, $\phi$ is a complex scalar, and $g_{S,P}$, $g^\prime_{S,P}$ are scalar ($S$) and pseudoscalar ($P$) couplings.  We assume $\chi_2$ and $\phi$ carry electric charge $Q_{\chi_2}|e|=Q_{\phi}|e|$ and are $SU(3)_C$-singlets.  We take $\chi_1$ to be Majorana.

\begin{figure}[t]
\includegraphics[scale=1]{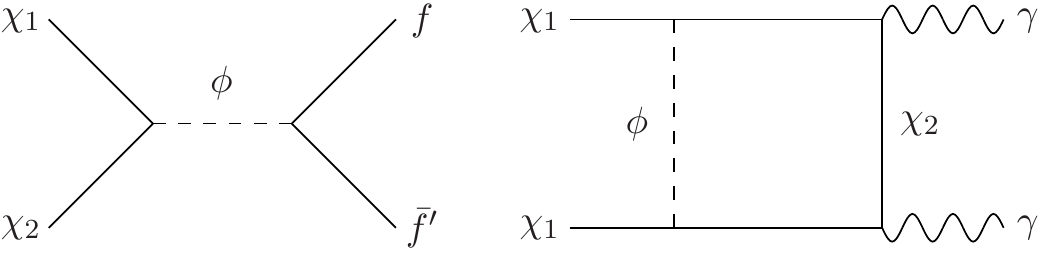}
\caption{Coannihilation $\chi_1 \chi_2 \to f \bar f^\prime$ ({\it Left}), where $\chi_1$ is DM, the coannihilating state $\chi_2$ and mediator $\phi$ carry electric charge, and $f,f^\prime$ are SM fermions.  $\chi_1 \chi_1 \to \gamma\gamma$ arises at one-loop ({\it Right}).}
\label{feyn3}
\end{figure}

The DM relic density is set by coannihilation $\chi_1 \chi_2 \to f \bar{f}^\prime$ and $\chi_1 \bar \chi_2 \to f^\prime \bar{f}$, shown in Fig.~\ref{feyn3}.  $\chi_2 \bar{\chi}_2 \to \gamma \gamma, f \bar f$ also occurs through gauge interactions.  The cross sections are
\begin{align}
\sigma_{12} v &= \sigma(\chi_1 \chi_2 \to f \bar{f}^\prime)v = \sigma  (\chi_1 \bar\chi_2 \to f^\prime \bar{f})v = \frac{|g_P|^2 \big(|g_S^\prime|^2 + |g_P^\prime|^2\big) (m_1 + m_2)^2}{8\pi \big((s-m_\phi^2)^2+m_\phi^2 \Gamma_\phi^2\big)} \label{coann}\\
\sigma_{22} v &= \sigma(\chi_2 \bar\chi_2 \to {\rm SM})v = \sigma(\chi_2 \bar\chi_2 \to \gamma \gamma)v + \sum_{f} \sigma(\chi_2 \bar\chi_2 \to f \bar f)v = \big(  Q^4_{\chi_2} + (20/3) Q^2_{\chi_2} \big) \frac{\alpha^2 \pi}{m_{2}^2} \; .
\end{align}
where the factor of $\sum_f N_c^f Q_f^2 = 20/3$ arises from the sum over all charged SM fermions except $t$, which is kinematically blocked.  The partial widths entering $\Gamma_\phi$ are
\begin{align}
\Gamma(\phi \to f \bar f^\prime) &= \frac{m_\phi}{8\pi} \big(|g_S^\prime|^2 + |g_P^\prime|^2\big) \\
\Gamma(\phi \to \chi_1 \chi_2 ) &= \frac{m_\phi}{8\pi} \Big( |g_S|^2 \Big(1 - \frac{(m_1 + m_2)^2}{m_\phi^2} \Big) + |g_P|^2 \Big( 1 - \frac{(m_1-m_2)^2}{m_\phi^2} \Big)\Big) \sqrt{1 - 2 \frac{m_1^2+m_2^2}{m_\phi^2} + \frac{(m_1^2 - m_2^2)^2}{m_\phi^4}} \; .
\end{align}
We work to lowest order in $v$, with the exception that we take $s=(m_1+m_2)^2(1+v^2/4)$ in Eq.~\eqref{coann} to properly account for a possible resonant enhancement~\cite{Griest:1990kh}; near the resonance, $\langle \sigma_{12} v\rangle$ must be computed numerically according to Eq.~\eqref{velavg}.  We neglect contributions to $\sigma_{22}$ from $Z$-exchange and $WW$ final states, which depend on the specific $SU(2)_L \times U(1)_Y$ quantum numbers of $\chi_2$ and $\phi$.  Finally, the total effective cross section is
\be
\langle \sigma_{\rm eff}v \rangle = 2 r_1 r_2\, \langle \sigma_{12} v\rangle + \frac{r_2^2}{2} \, \langle \sigma_{22} v\rangle 
\ee
where $r_{1,2}$ are defined as in Eq.~\eqref{rdef} with $g_1 = 2$, $g_2 = 4$. The relic density is given by Eq.~\eqref{relic}.

\begin{figure}
\includegraphics[scale=.7]{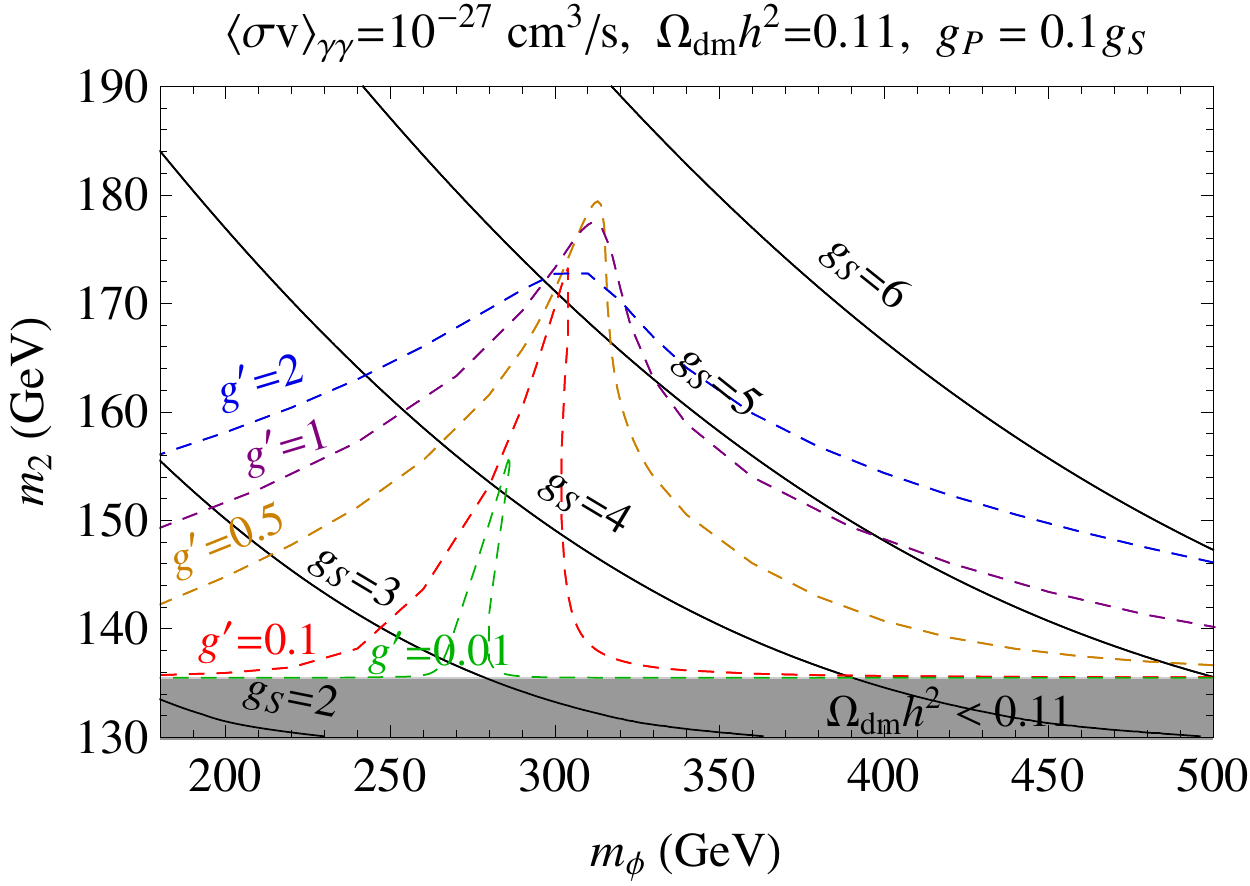}~
\includegraphics[scale=.7]{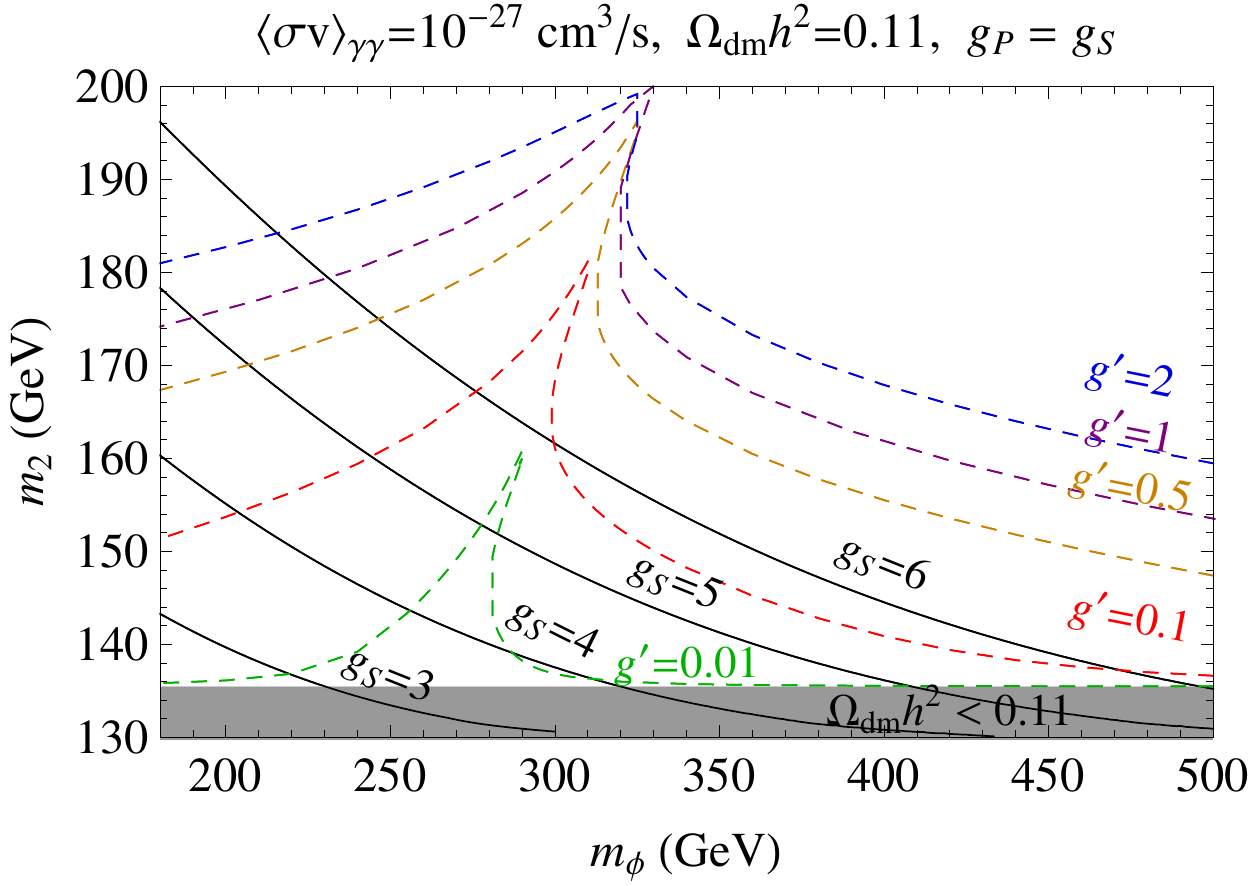}
\caption{Solid contours show masses $(m_2,m_\phi)$ and couplings $g_{S,P}$ for $\langle\sigma v\rangle_{\gamma\gamma} = 10^{-27} \, \cms$, with $g_P=0.1 g_S$ ({\it Left}) and $g_P=g_S$ ({\it Right}) and $Q_{\chi_2} = 1$.  Dashed contours show $\Omega_{\rm dm}h^2 = 0.11$, for different values of $g^\prime \equiv \sqrt{|g^\prime_S|^2+|g^\prime_P|^2}$.  Gray region is excluded by $\Omega_{\rm dm}h^2 < 0.11$.}
 \label{fig:coann}
 \end{figure}

DM can annihilate directly into $\gamma\gamma$ at one-loop, shown in Fig.~\ref{feyn3}, generating the $\gamma$ line signal.  The cross section, given in Ref.~\cite{Bergstrom:1997fh,Bern:1997ng,Ullio:1997ke}, is
\be
\langle\sigma v\rangle_{\gamma\gamma} = \sigma  (\chi_1 \chi_1 \to \gamma \gamma)v = \frac{\alpha^2 Q^4_{\chi_2}  m_1^2}{64 \pi^3 m_\phi^4}\,  \big(F_+ |g_S|^2 + F_- |g_P|^2 \big)^2 \, .
\ee
We have defined
\be
F_\pm \equiv \frac{1}{a} \left[ \frac{a \pm \sqrt{a b}}{1+a-b} I_1(a,b) + \frac{1}{1-b} I_2(a,b) + \left(\frac{2b \pm 2\sqrt{a b}}{1+a-b} - \frac{b}{1-b} \right) I_3(a,b) \right] \, .
\ee
where $a \equiv m_{1}^2/m_\phi^2$, $b \equiv m_{2}^2/m_\phi^2$, and the functions $I_n(a,b)$ are defined in \cite{Bergstrom:1997fh}.  In the $m_{\phi} \gg m_{1,2} \gg \Delta m \equiv m_2 - m_1$ limit, we have $F_+ \approx (2-\pi^2)$ and $F_- \approx 2$; however, for $m_\phi \sim m_{1,2}$, these approximations overestimate the $\gamma\gamma$ rate and we use the exact expression in our analysis.  Also, we expect the rates for $\chi_1 \chi_1 \to Z Z, Z\gamma$ to be comparable, although the exact prediction depends on the $SU(2)_L \times U(1)_Y$ quantum numbers of $\chi_2$ and $\phi$.

In Fig.~\ref{fig:coann}, we present numerical results for this model.
\begin{itemize}
\item The solid curves show mass contours for $\langle\sigma v\rangle_{\gamma\gamma}  = 10^{-27} \, \cms$, for fixed $m_1 = 130$ GeV and for different couplings $g_S$, with $g_P = 0.1\,g_S$ (left panel) and $g_P = g_S$ (right panel).  The $\gamma$ line signal requires $g_S \gtrsim O(1)$ and $m_2, m_\phi \gtrsim m_1$.
\item The dashed contours show parameters giving the DM relic density $\Omega_{\rm dm} h^2 = 0.11$, for different values of the SM fermion coupling $g^\prime \equiv \sqrt{|g_S^\prime|^2 + |g_P^\prime|^2}$, with $m_1 = 130$ GeV and $g_{S,P}$ fixed by $\langle\sigma v\rangle_{\gamma\gamma} $.  There is a clear resonance for $m_\phi \approx m_1 + m_2$, with smaller values of $g^\prime$ and larger $\Delta m$ allowed.    (The width $\Gamma_\phi$ is computed as a function of the given parameters.)
\item The gray region is excluded by $\Omega_{\rm dm} h^2 < 0.11$. For $\Delta m \lesssim 5$ GeV, $\chi_2 \bar{\chi}_2$ annihilation is not sufficiently Boltzmann suppressed, depleting $\chi_1$ provided $\chi_1$ and $\chi_2$ are in chemical equilibrium.  (This holds for $g_{S,P} \sim 1$, $g^\prime \gg 10^{-7}$.)  
\end{itemize}
Taking $m_2 \approx 135~\gev$ (corresponding to the edge of the gray region) gives $\Omega_{\rm dm} h^2 = 0.11$ in a large region of parameter space ($10^{-7} \ll  g^\prime \ll 10^{-1}$, off-resonance) with little dependence on the other new physics parameters, since the relic density is set through electromagnetic interactions.  That is, the new physics particles need not have large couplings to SM states, aside from their electromagnetic couplings.  In any case, this coannihilation model presents a viable framework for explaining the DM relic density with an enhanced $\gamma$ line signal.

\section{Forbidden channels}
\label{sec:forbidden}

The second exception occurs when {\em all} the virtual charged particles generating the DM coupling to photons have a slightly larger mass than the DM.  Although the coupling between DM and the charged particles has to be strong to overcome the loop-suppression factor, the annihilation cross section to charged particles at tree-level is suppressed kinematically.  During freeze-out, DM is non-relativistic and its typical velocity is $\sim0.3~c$. If the charged particles have masses not far from the DM mass, annihilation to the charged particles can still proceed in the early Universe, albeit less efficiently.  As a result, one is able to obtain the correct relic density despite the large couplings needed to generate a photon line.  On the other hand, DM has a typical velocity $\sim10^{-3}~c$ in the halo today so that the direct annihilation to the charged particles is kinematically forbidden, evading constraints from continuum photons.  In Ref.~\cite{Jackson:2009kg}, this mechanism was used to generate enhanced DM annihilation to $\gamma Z$ and $\gamma h$, with the forbidden particle as the $t$ quark.  Here, we investigate a different model with enhanced annihilation to $\gamma\gamma$, and we compute the required mass splitting between the forbidden states and DM to obtain the correct relic density and the Fermi line signal simultaneously.

\begin{figure}[t]
\includegraphics[scale=1]{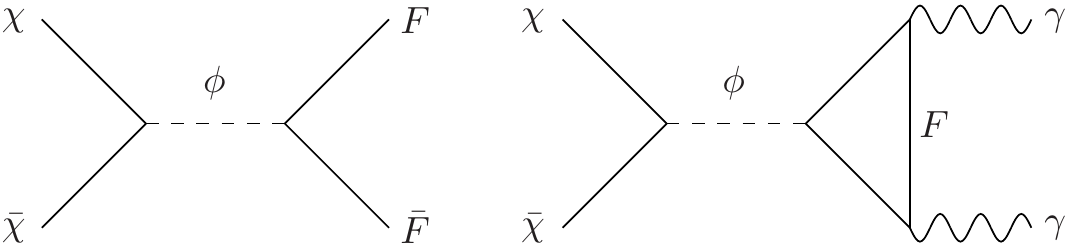}
\caption{Tree-level dark matter annihilation to heavy fermions in the forbidden case ({\it Left}). $\chi\chi\rightarrow\gamma\gamma$ at one-loop ({\it Right}).  }
\label{feyn4}
\end{figure}

We proceed to estimate the relic density through annihilation to the charged particle pairs, $\bar{\chi}\chi\rightarrow F\bar{F}$, where we use $F$ to denote charged fermions heavier than DM. 
We begin by reviewing the discussion of \cite{Griest:1990kh}.  Since the velocity of the final-state particles is small, it is convenient to write the annihilation cross section in the form $(\sigma v)=(a+bv^2)v_2$, where $v$ is the relative velocity of the initial-state particles, $v_2$ is the velocity of the final-state particles in the center of mass frame, and $a$ and $b$ characterize the $s$-wave and $p$-wave contributions to the annihilation cross section respectively as usual.\footnote{The reader should not be confused with the mass ratios $a,b$ defined in Sec.~\ref{sec:coann}. Here, $a,b$ refer to $s$- and $p$-wave cross sections only.} Note $v_2$ must present in the annihilation cross section because it is from the phase space of the final-state particles. Energy and momentum conservation require
\be
v_2=\sqrt{1-\left(\frac{m_F}{m_\chi}\right)^2+\left(\frac{m_F}{m_\chi}\right)^2\frac{v^2}{4}}.
\ee
The important step in computing the relic density for the forbidden case is to evaluate the thermally-averaged annihilation cross section, given by
\be
\left<\sigma v\right>=\left<(a+b v^2)v_2\right>=\frac{x^{3/2}}{2\pi^{1/2}}\int^\infty_{2\mu_-} v_2(a+b v^2)v^2e^{-v^2x/4}dv,
\label{eq:forbi}
\ee
where $\mu_-=(1-m^2_\chi/m^2_F)^{1/2}$. Note $2\mu_-$ is the minimal velocity to activate the annihilation. The integral of Eq.~(\ref{eq:forbi}) can be performed numerically. In the case of an $s$-wave cross section off resonance, an analytical result is possible 
\begin{eqnarray}
\left<a v_2\right>=a\frac{\mu^2_- z x^{1/2}}{\pi^{1/2}}e^{-\mu^2_-x/2}K_1(\mu^2_-x/2),
\label{eq:forbi2}
\end{eqnarray}
where $z=m_F/m_\chi$ and $K_1$ is the modified Bessel function~\cite{Griest:1990kh}.
The relic DM density of $\chi$ is
\be
\Omega_{\rm dm} h^2=\frac{1.07\times10^9~\gev^{-1}}{g^{1/2}_*m_{\rm Pl}\left[\int^\infty_{x_f}x^{-2}\left<\sigma v\right>dx\right]} \, ,
\ee
where as usual the freeze-out temperature is $x_f=\ln\big(0.038gm_\chi m_{\rm Pl}\left<\sigma v\right>/\sqrt{g_*x_f}\big)$.

Having reviewed the relic density calculation in the forbidden case, we consider a concrete example. We assume that the DM $\chi$ is a Majorana fermion and it couples to charged fermions through a pseudoscalar mediator.  Pseudoscalar couplings are needed for a scalar mediator case to obtain an $s$-wave annihilation to $\gamma \gamma$. The interaction Lagrangian is given by
\be
\mathscr{L}_{\rm int}=\frac{ig_\chi}{2}\phi\bar{\chi}\gamma_5\chi+ig_F\phi\bar{F}\gamma_5 F,
\ee
where $F$ is a charged fermion. Since we need $m_F\gtrsim 130~{\rm GeV}$, the only possible candidate for $F$ among SM fermions is the top quark. If $F$ carries $SU(2)_L$ quantum numbers, there are comparable annihilation cross sections to $ZZ$ and $Z\gamma$, while if $F$ carries only hypercharge, the $\gamma\gamma$ channel will dominate as discussed for the dipolar DM case. In the limit $m_F\geq m_\chi$, the annihilation cross section to photons through an $F \bar F$ loop is 
\be
(\sigma v)_{\gamma\gamma}=\sigma(\chi\chi\rightarrow\gamma\gamma)v=\frac{1}{4\pi^3}\frac{\alpha^2g^2_\chi g^2_F Q^4_F c^2_F m^2_F}{(s-m^2_\phi)^2+m^2_\phi\Gamma^2_\phi}\left[\arctan\left(\frac{1}{\sqrt{m^2_F/m^2_\chi-1}}\right)\right]^4,
\label{eq:forgg}
\ee
where $\Gamma_\phi$ is the total decay width of $\phi$, $Q_F$ is the electric charge of $F$ in units of $|e|$ and $c_F$ is its color quantum number. In this model, $\Gamma_\phi$ is a sum of the following decay widths 
\be
\Gamma(\phi\rightarrow F\bar{F})=\frac{m_\phi}{8\pi}g^2_F\sqrt{1-\frac{4m^2_F}{m^2_\phi}} , \;\;
\Gamma(\phi\rightarrow \chi\chi) =\frac{m_\phi}{16\pi}g^2_\chi\sqrt{1-\frac{4m^2_\chi}{m^2_\phi}} , \;\;
\Gamma(\phi\rightarrow \gamma\gamma)=\frac{m^3_\phi\alpha^2Q^4_F}{256\pi^3 m^2_F}g^2_F \big|A^A_{1/2}\big(m^2/4m^2_F\big) \big|^2, \label{eq:forwidth}
\ee
where the function $A^A_{1/2}(\tau)$ is given by $A^A_{1/2}(\tau)\equiv2\tau^{-1}f(\tau)$ with 
\be
f(\tau)=\left\{ \begin{array}{ll} \big(\arcsin\sqrt{\tau}\big)^2 & {\rm for} \; \tau\leq1 \\
-\frac{1}{4}\big(\log\frac{1+\sqrt{1-\tau^{-1}}}{1-\sqrt{1-\tau^-1}}-i\pi\big)^2 &{\rm for}\;\tau>1 \end{array} \right.. 
\ee
In our numerical study, $\Gamma_\phi$ is computed as a function of the given parameters according to Eq.~(\ref{eq:forwidth}).

Since $\chi\chi\rightarrow\gamma\gamma$ is dominated by the $s$-wave process, the thermally-averaged annihilation cross section $\langle\sigma v\rangle_{\gamma\gamma}$ equals $(\sigma v)_{\gamma\gamma}$. The annihilation cross section to $F\bar{F}$ is
\be
(\sigma v)_{F\bar{F}}=\sigma(\chi\chi\rightarrow F\bar{F})v=\frac{1}{2\pi}\frac{g^2_\chi g^2_F c_F m^2_\chi}{(s-m^2_\phi)^2+m^2_\phi\Gamma^2_\phi}v_2.
\label{eq:ffbar}
\ee
The relevant diagrams are shown in Fig.~\ref{feyn4}.

\begin{figure}
\includegraphics[scale=0.9]{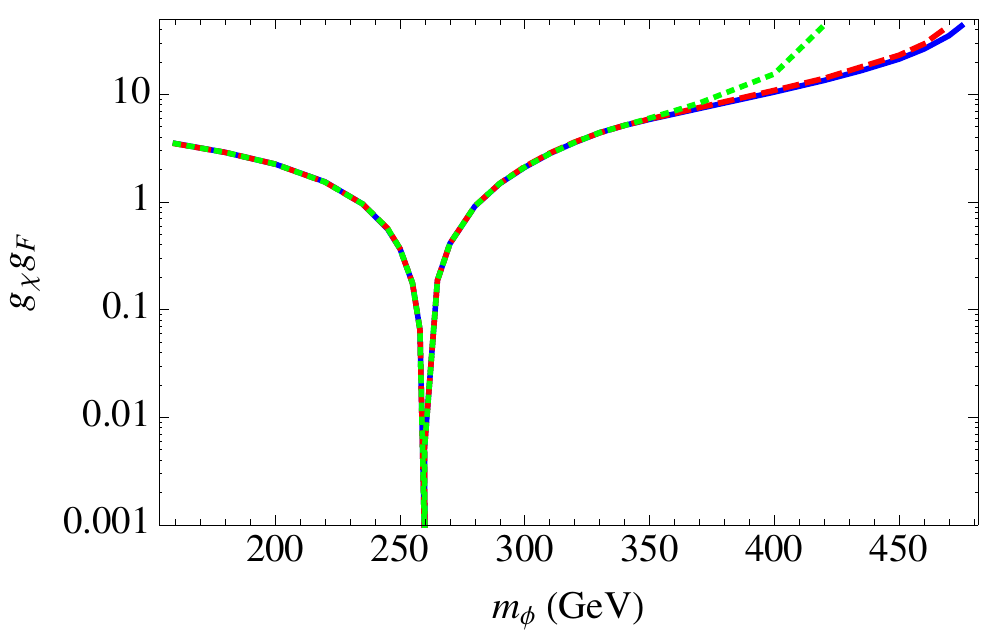}
\includegraphics[scale=0.85]{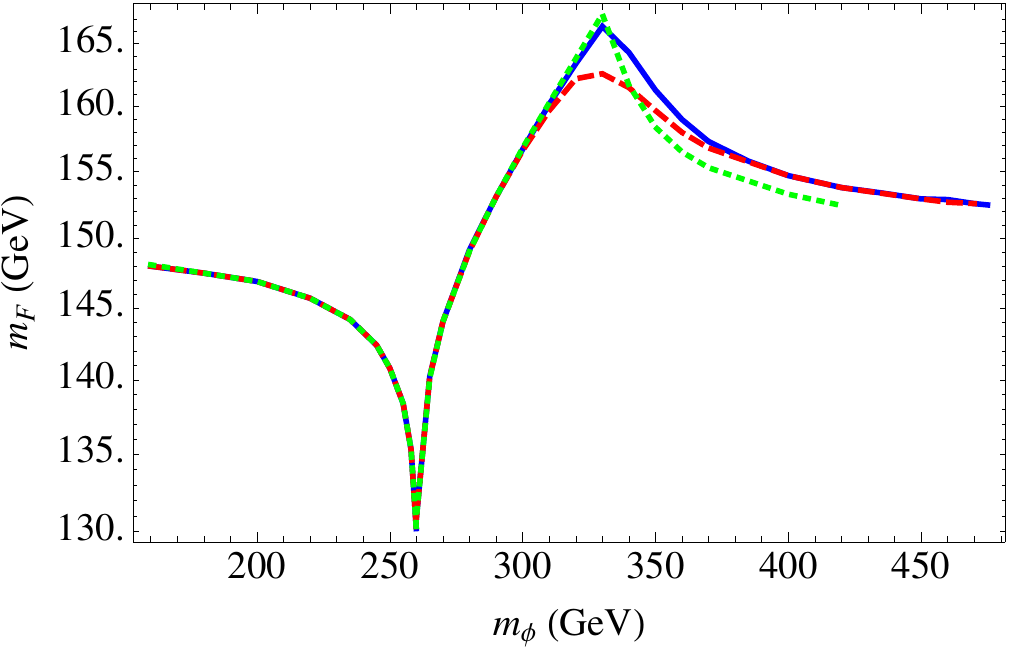}
 \caption{Contours show the coupling $g_\chi g_F$ ({\it Left}) and the heavy charged particle mass $m_F$ ({\it Right}) required for $\langle\sigma v\rangle_{\gamma\gamma}=10^{-27}~{\rm cm^3/s}$ and $\Omega_\chi h^2=0.11$ as a function of the mediator mass $m_\phi$ in the forbidden case. We take $m_\chi=130~\gev$, $Q_F=c_F=1$, $g_\chi=0.5 g_F$ (dotted), $g_\chi=g_F$ (solid), and $g_\chi=2g_F$ (dashed). All contours stop when $g_\chi g_F\sim{\cal O}(40)$.}
 \label{fig:for}
 \end{figure}

Using Eq.~(\ref{eq:forgg}), we can estimate the magnitudes of coupling constants required to generate the photon line signal. When $m_\phi$ is far from $2m_\chi$, a large coupling constant $g_\chi g_F\sim{\cal O}(4-10)$ is required for $\langle\sigma v\rangle_{\gamma\gamma}=10^{-27}~\cms$, depending on the mediator mass. While $m_\phi\approx2m_\chi$,  the line signal can be enhanced dramatically. In this resonance limit, the annihilation cross section to photons can be approximated as 
\be
(\sigma v)_{\gamma\gamma}\approx10^{-27}~{\rm cm^3/s}\left(\frac{g^2_\chi g^2_F Q^4_F c^2_F}{2\times10^{-4}}\right)\left(\frac{m_F}{130~{\rm GeV}}\right)^2\left(\frac{260~{\rm GeV}}{m_\phi}\right)^2\left(\frac{1 \mbox{ GeV}}{\Gamma_\phi}\right)^2.
\ee
Therefore, the line signal can be enhanced significantly in the resonance case and the required coupling constants can be much less than ${\cal O}(1)$. 

Next, we discuss the thermal relic density for $\chi$. Since $\chi\chi\rightarrow F{\bar F}$ is dominated by the $s$-wave process, we only keep the $a$ term in the expansion of $(\sigma v)=(a+bv^2)v_2$, which is given by
\begin{eqnarray}
a=\frac{1}{2\pi}\frac{g^2_\chi g^2_F c_F m^2_\chi}{(s-m^2_\phi)^2+m^2_\phi\Gamma^2_\phi},
\end{eqnarray}
where $s=4m^2_\chi/(1-v^2/4)$ with a minimal $v$ as $2(1-m^2_\chi/m^2_F)^{1/2}$. In our numerical work, we take the thermal average on the whole annihilation cross section $av_2$ as in Eq.~(\ref{eq:forbi}).  This is important to calculate the relic density near resonance. We also have checked that one may take $s=4m^2_F$ and use Eq.~(\ref{eq:forbi2}) directly if it is off resonance.

To see how we can enhance $\gamma\gamma$ signals and obtain the DM density simultaneously in the forbidden case, it is suggestive to check the ratio of $(\sigma v)_{\gamma\gamma}$ to $(\sigma v)_{F\bar{F}}$. Taking $Q_F=c_F=1$ and $m_F\gtrsim m_\chi$, we have
\be
\frac{(\sigma v)_{\gamma\gamma}}{(\sigma v)_{F\bar{F}}}\approx 2\times10^{-5} \times \frac{1}{v_2}\frac{(4m^2_F-m^2_\phi)^2+m^2_\phi\Gamma^2_\phi}{(4m^2_\chi-m^2_\phi)^2+m^2_\phi\Gamma^2_\phi}.
\label{eq:ggff}
\ee
We see that there are two effects can overcome the loop suppression factor and boost $(\sigma v)_{\gamma\gamma}$ with respect to $(\sigma v)_{F\bar{F}}$.  The first is the phase space factor $v_2$; for $m_F \gtrsim m_\chi$, we have $v_2 \ll 1$. 
The second boost factor is from a resonance effect.  Since $F\bar F$ annihilation occurs at $s \approx 4m_F^2$, while $\gamma\gamma$ annihilation occurs at $s\approx 4m_\chi^2$, the latter can be enhanced by a pole at $m_\phi\approx2m_\chi$.  Both effects rely on forbidden channels.  If $m_F \ll m_\chi$, then $v_2 \sim 1$ and both $F\bar F$ and $\gamma\gamma$ annihilation have the same resonant enhancement because they have a same pole at $m_\phi = 2m_\chi$.
Therefore, a successful implementation of these enhancements relies on the mass gap between $F$ and $\chi$.


We present our numerical results for the forbidden case on two complementary panels of Fig.~\ref{fig:for}. In Fig.~\ref{fig:for} ({\it Left}), we show $g_\chi g_F$ required for the DM relic density and $\langle\sigma v\rangle_{\gamma\gamma}=10^{-27}~{\rm cm^3/s}$ as a function of $m_\phi$. For each point along the contour, the value of $m_F$ is given in Fig.~\ref{fig:for} ({\it Right}). When $m_\phi\approx 2m_\chi$,  $\chi\chi\rightarrow\gamma\gamma$ is enhanced and a small coupling constant is needed to generate the Fermi line signal. In this case, a relatively small $m_F$ is required to suppress annihilation to $F\bar{F}$. It is interesting to note that $m_F$ has to be very close to $130~\gev$ to obtain the correct relic density when $\gamma\gamma$ is enhanced maximally. On the other hand, $(\sigma v)_{F\bar{F}}$ is on resonance during freeze-out for $m_\phi\approx2m_F$. Therefore, one needs larger $m_F$ to suppress the boosted annihilation. We also can see that  the numerical result only has a mild dependence on the relative size of $g_\chi$ and $g_F$. This is because the value of each coupling enters the calculation individually only through $\Gamma_\phi$, which is only important near the resonance. The dependence is negligible for small $m_\phi$, where the width is very narrow and it does not play a role. For $m_\phi\gtrsim 300-350~\gev$, the effect is more noticeable since more decay channels become kinematically accessible. In this model, the preferred value of $m_F$ is $\sim130-165~\gev$ depending on parameters. With such heavy charged particles, it is clear that $\chi\chi\rightarrow F\bar{F}$ is forbidden kinematically in the galaxy today, and the model evades the continuum photon constraint.

\section{Asymmetric Dark Matter}
\label{sec:adm}

Asymmetric DM (ADM)~\cite{Kaplan:2009ag} provides a third exception for reconciling an enhanced $\gamma$ line signal with the observed relic density.\footnote{For early ADM works, see~\cite{Nussinov:1985xr,Barr:1990ca,Barr:1991qn,Kaplan:1991ah,Berezhiani:1995yi}; for more recent works, see \cite{Davoudiasl:2012uw} and Refs.~therein.}  We assume that DM $\chi$ is a complex state carrying a $U(1)_\chi$ conserved charge, and that a nonzero $\chi$ chemical potential arises sometime before the freeze-out epoch, generating an asymmetry of $\chi$ over its antiparticle $\chi^\dagger$.
In ADM freeze-out, the $\chi \chi^\dagger$ annihilation cross section can be much larger than $\sim6 \times 10^{-26} \, \cms$ required for symmetric freeze-out.  In this case, $\chi \chi^\dagger$ annihilation is quenched once $\chi^\dagger$ is depleted, and the relic $\chi$ density is determined by the primordial asymmetry.  This is similar in spirit to coannihilation, where the coannihilating state $\chi^\dagger$ is suppressed by a chemical potential, rather than a mass splitting.  

DM annihilation can occur in the Universe today if the $\chi$ asymmetry is washed out after freeze-out through $\chi \leftrightarrow \chi^\dagger$ oscillations~\cite{Cohen:2009fz,Buckley:2011ye,Cirelli:2011ac,Tulin:2012re}.  Particle-antiparticle oscillations are generic in a wide class of ADM models where, unless $U(1)_\chi$ descends from a gauge symmetry, one expects $U(1)_\chi$-breaking mass terms to arise, e.g., through Planck-suppressed operators.  In this case, $\chi$ and $\chi^\dagger$ are no longer mass eigenstates, and oscillations commence once the mass splitting between the real components of $\chi$ is comparable to the Hubble expansion rate.

\begin{figure}[t]
\includegraphics[scale=1]{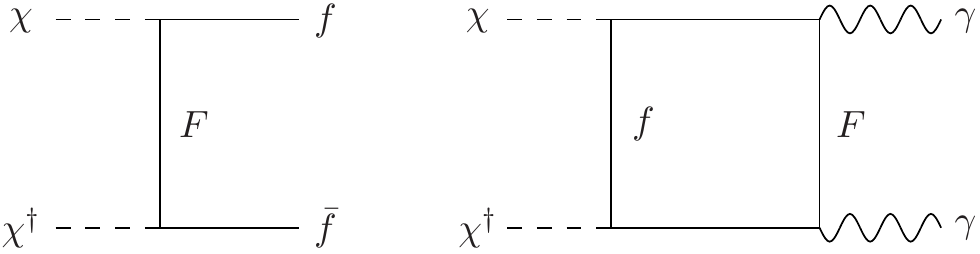}
\caption{Scalar DM $\chi$ annihilating to SM fermions $f \bar f$ ({\it Left}) and $\gamma\gamma$ ({\it Right}), where $F$ is a new massive charged fermion.}
\label{feynadm}
\end{figure}

We consider a model where $\chi$ is a complex scalar with an interaction
\beq
{\mathscr L}_{\rm int} = \chi \bar F (g_L P_L + g_R P_R) f + \mbox{ h.c.} \, , \label{Lint}
\eeq 
where $g_{L,R}$ are couplings, $f$ is a SM fermion, and $F$ is a new massive fermion carrying $U(1)_\chi$ with mass $m_F>m_\chi$.  We assume $f,F$ carry electric charge $Q_f |e| = Q_F |e|$.  DM directly annihilates to $f \bar f$ at tree-level and to $\gamma\gamma$ at one-loop, shown in Fig.~\ref{feynadm}.  Since one expects the former to be enhanced over the latter by $\mathcal{O}(\pi^2/\alpha^2)$, we must address how this model can generate the observed $\gamma$ line while avoiding $\gamma$ continuum constraints.

The case of scalar DM provides a natural mechanism to suppress annihilation to $f \bar f$, thereby evading the $\gamma$ continuum constraint.  If $\chi$ couples chirally, $\chi\chi^\dagger \to f \bar f$ is $p$-wave or chirality-suppressed as a consequence of angular momentum conservation.  Taking, {\it e.g.}, $g_L=0$, we have
\be \label{chiralsigma}
\sigma(\chi\chi^\dagger \to f \bar f) v \approx \frac{|g_R|^4 ( 3 m_f^2  + m_\chi^2 v^2 )}{48\pi (m_\chi^2 + m_F^2)^2} \; ,
\ee
keeping only the leading terms in $v^2$ or $m_f^2$.  On the other hand, if $g_{L}\sim g_R \ne 0$, the leading contribution is $s$-wave and is not chirality-suppressed:
\be \label{ffbar}
\sigma(\chi\chi^\dagger \to f \bar f) v \approx \frac{|g_L|^2 |g_R|^2 m_F^2}{4 \pi(m_\chi^2 + m_F^2)^2} \; .
\ee
For example, in the case of $f=\tau$, the annihilation rate in the galactic halo today $(v \sim 10^{-3})$ is
\be
\sigma(\chi\chi^\dagger \to \tau \bar \tau) v \approx \left\{ \begin{array}{ll} 10^{-23} \, \cms \times |g_L|^2 |g_R|^2 & {\rm for} \; g_L \sim g_R \\
6\times 10^{-28} \, \cms \times |g_R|^4 & {\rm for} \;  g_L = 0 \end{array} \right. \, , 
\ee
taking $m_F \sim m_\chi = 130$ GeV.  Clearly, $\mathcal{O}(1)$ chiral couplings are consistent with $\gamma$ continuum constraints, while nonchiral couplings are much more strongly constrained.

\begin{figure}[t]
\includegraphics[scale=.68]{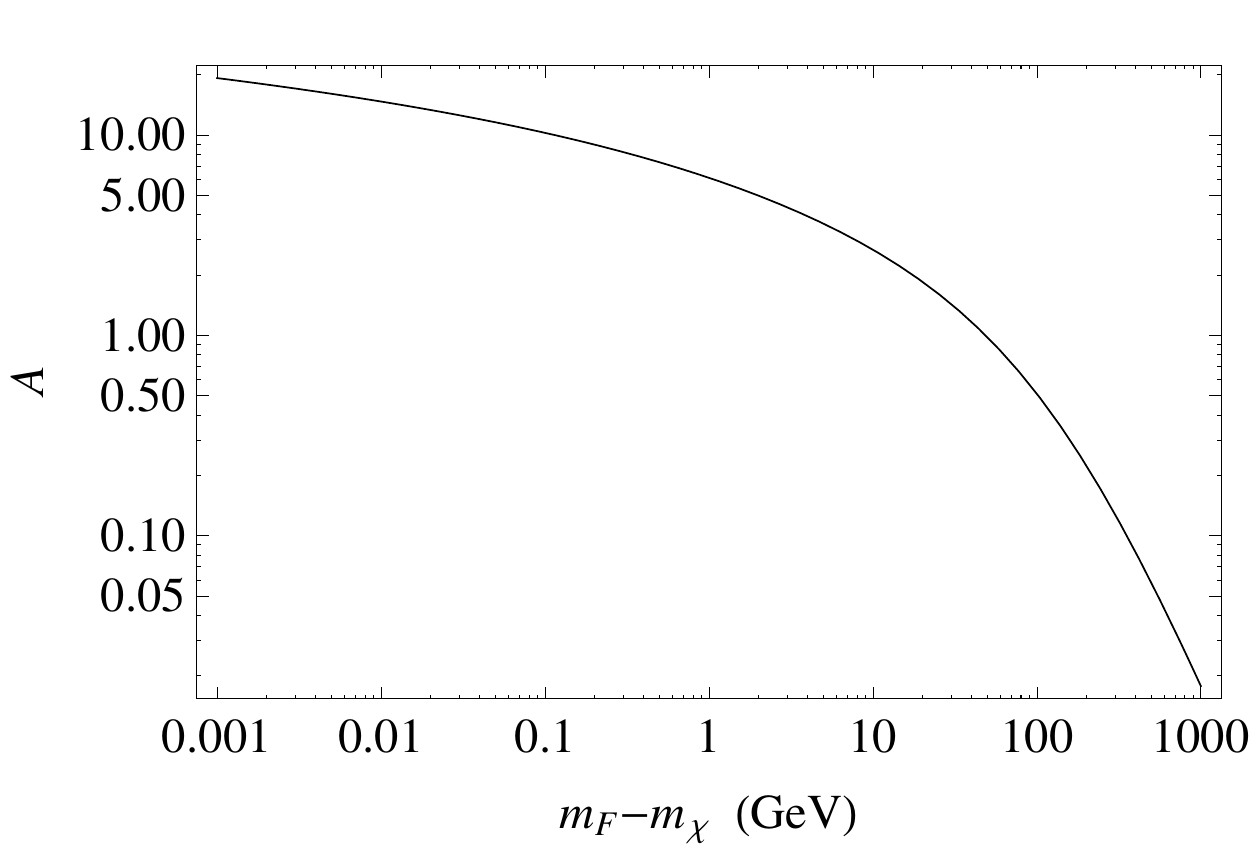} ~\includegraphics[scale=.68]{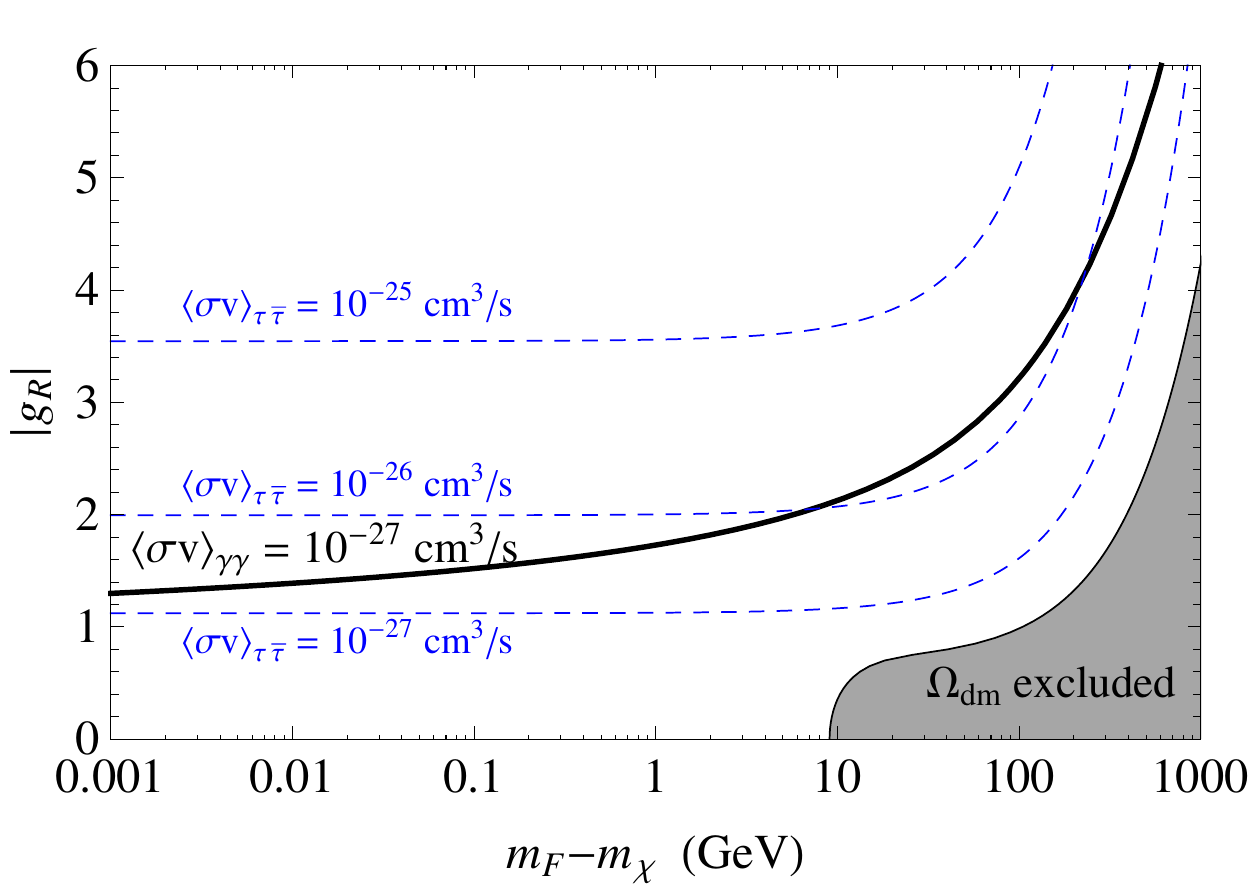}
\caption{{\it Left}: Matrix element $\mathcal{A}$ as a function of mass splitting $m_F - m_\chi$, for $m_\chi = 130$ GeV and $m_f=0$. {\it Right}: Solid contour shows coupling $|g_R|$ and mass splitting $m_F-m_\chi$ required for $\langle \sigma v\rangle_{\gamma\gamma} = 10^{-27} \, \cms$, for $m_\chi = 130$ GeV and $g_L=0$; dashed blue contours show $\chi \chi^\dagger \to \tau \bar\tau$ cross sections for $f=\tau$ case.  Shaded region is excluded by ADM relic density considerations (see text).}
\label{admplots}
\end{figure}

The cross section for $\chi \chi^\dagger \to \gamma \gamma$ is given by
\begin{equation}
\langle \sigma v \rangle_{\gamma \gamma} = \frac{\alpha^2 Q_f^4 (|g_L|^2+|g_R|^2)^2}{64 \pi^3 m_\chi^2} |{\mathcal A}|^2 \approx 2 \times 10^{-29} \, \cms \times Q_f^4 (|g_L|^2 + |g_R|^2)^2 |\mathcal{A}|^2 \, .
\label{boxresonance}
\end{equation}
The matrix element $\mathcal{A}$, computed in Ref.~\cite{Bertone:2009cb} for $m_f = 0$, can be expressed as
\begin{align}
\mathcal{A} = 2 - 2 \log\big( 1-\tau \big) - 2\tau^{-1} \, \arcsin^2\big(\sqrt{\tau}\big)    \,    ,\label {resbox}
\end{align}
where $\tau=m_\chi^2/m_F^2$.  The numerical value of $\mathcal{A}$ is shown in Fig.~\ref{admplots} ({\it Left}). Although $\mathcal{A}$ diverges logarithmically for $\tau \to 1$, we expect the analytical formula to break down when $\tau \approx 1- m_f^2/m_\chi^2$ since $\mathcal{O}(m_f^2)$ terms have been neglected. 

In Fig.~\ref{admplots} ({\it Right}), we show numerical results for $\chi \chi^\dagger$ annihilation cross sections for $m_\chi = 130$ GeV and $g_L=0$.  The solid contour shows the coupling $g_R$ and mass splitting $m_F-m_\chi$ required for $\langle \sigma v \rangle_{\gamma\gamma} = 10^{-27} \, \cms$ to explain the Fermi $\gamma$ line signal.  The required parameters are easily consistent with $\gamma$ continuum constraints on $\chi \chi^\dagger \to f \bar f$.  For example, taking $f=\tau$, the dashed blue contours show the $\chi \chi^\dagger \to \tau \bar \tau$ cross section, easily consistent with present constraints~\cite{GeringerSameth:2011iw,Ackermann:2011wa, Ackermann:2012qk}. Note the cases with $f=e,\mu$ are even less constrained by continuum constraints due to the chirality suppression.

Lastly, we discuss constraints from DM relic density considerations.  ADM freeze-out in the early Universe requires a large annihilation cross section $\langle \sigma v \rangle \gtrsim 6 \times 10^{-26} \, \cms$ to deplete the symmetric $\chi$ density, leaving behind the residual asymmetric component.  Although $\chi \chi^\dagger \to f \bar f$ is suppressed today, annihilation is greatly enhanced in the early Universe in two ways: (i) the DM velocity during freeze-out is $v \sim 0.3$, enhancing the $p$-wave term in Eq.~\eqref{chiralsigma}, and (ii) for $m_F - m_\chi \lesssim 10$ GeV, coannihilation becomes important.  The total effective annihilation cross section is (see Sec.~\ref{sec:coann})
\be
\langle \sigma_{\rm eff} v \rangle = r_{\chi}^2 \langle \sigma(\chi \chi^\dagger \to f \bar f)v\rangle + 2 r_{\chi} r_F \langle \sigma(\chi F \to \gamma \bar f)v\rangle + r_F^2 \langle \sigma(F \bar F \to {\rm SM})v\rangle 
\ee
with coannihilation cross sections
\be
\sigma(\chi F \to \gamma \bar f)v = \frac{\alpha Q_F^2 (|g_L|^2+|g_R|^2) m_\chi}{8 m_F^2(m_\chi+m_F)} \; , \qquad 
\sigma(F \bar F \to {\rm SM})v \approx \big(  Q^4_{F} + (20/3) Q^2_{F} \big) \frac{\alpha^2 \pi}{m_{F}^2} \, ,
\ee
where for $F \bar F$ annihilation into SM particles we include only the dominant electromagnetic terms, summing over $\gamma\gamma$ and all fermions except $t$.  Considering the case where $f = \tau$, $g_L=0$, and $m_\chi=130$ GeV, the gray region in Fig.~\ref{admplots} is excluded by requiring $\langle \sigma_{\rm eff} v \rangle > 6 \times 10^{-26}$ at $x_f=25$.  That is, the $\gamma$ line signal is fully consistent with ADM freeze-out.  Parameters where {\it symmetric} DM gives the correct relic density correspond to the border of the gray and white regions, and therefore DM must be asymmetric in this model to explain the $\gamma$ line signal.

In addition, we require that $\chi \leftrightarrow \chi^\dagger$ oscillations begin during or after the freeze-out epoch to wash out the DM asymmetry, giving rise to observable annihilation signals today.  Therefore, the $U(1)_\chi$-breaking mass splitting should be less than $H(T_f) \sim 4\times10^{-8}~{\rm eV}$; otherwise, DM is symmetric since the asymmetry is erased before freeze-out.  If  oscillations occur much later than freeze-out, the DM relic density today is fixed by an initial asymmetry of $\mathcal{O}(3.5\%)$ of the baryon asymmetry. On the other hand, if oscillations begin during or soon after freeze-out, residual annihilation occurs and  larger DM asymmetries are required to give the observed $\Omega_{\rm dm}$~\cite{Cirelli:2011ac,Tulin:2012re}.

\section{Conclusions}
\label{sec:conclusions}

Recent analyses of Fermi LAT data have found evidence for a $\gamma$ line signal from the galactic center at $E_\gamma \approx 130$ GeV, with potentially a second line around $111$ GeV.  If these signals originate from DM annihilation, the required annihilation cross section to $\gamma\gamma$ is $\langle \sigma v \rangle_{\gamma\gamma} \approx 10^{-27} \: \cms$, relatively larger than in generic WIMP models.  To explain an enhanced $\gamma\gamma$ rate, one requires large DM couplings to light charged states, {\it e.g.}, fermion pairs $f \bar f$ or $W W$, generating a $\gamma\gamma$ coupling at one-loop.  One expects tree-level annihilation to $f \bar f$ and $WW$ to be enhanced over $\gamma\gamma$ by $\mathcal{O}(\pi^2/\alpha^2)$. Therefore, a WIMP interpretation of the Fermi line signal faces two obstacles: (i) annihilation to charged SM particles in the early Universe is too large to explain the DM relic density, leading to excessive DM depletion during freeze-out, and (ii)
annihilation to charged SM particles in the galactic halo today is too large, in conflict with Fermi LAT constraints on the continuum $\gamma$ spectrum produced by final state emission.

In this work, we have emphasized three exceptions to these obstacles.  For each case, annihilation to SM particles in the early Universe and in the halo today is suppressed, allowing for large DM couplings and an enhanced $\gamma\gamma$ rate, while giving the correct relic density and satisfying $\gamma$ continuum constraints for DM mass $m_\chi \approx 130$ GeV.  The three exceptions are:
\begin{itemize}
\item {\it Coannihilation:} The relic density is set by coannihilation $\chi_1 \chi_2 \to f \bar f$. An $\mathcal{O}(10 \:{\rm GeV})$ mass splitting between DM $\chi_1$ and the nearby state $\chi_2$ gives the right suppression to $f \bar f$ to explain both the relic density and $\gamma\gamma$ rate.  Annihilation to $f \bar f$ is absent in the halo since $\chi_2$ is not populated today.  One natural example is a DM transition magnetic dipole interaction.  We also considered a simple model where DM coannihilates with a state carrying electric charge.
\item {\it Forbidden channels:} DM annihilates $\chi \chi \to F \bar F$, where $F$ is a charged state slightly heavier than $\chi$.   Annihilation to $F \bar F$ is kinematically forbidden in the halo today, but occurs in the early Universe due to the higher DM velocity.   We obtain the correct relic density for $m_F \sim 150$ GeV.
\item {\it Asymmetric DM:} Due to a primordial $\chi$ asymmetry, DM annihilation $\chi \chi^\dagger \to f \bar f$ becomes suppressed in the early Universe when the symmetric $\chi$,$\chi^\dagger$ density is depleted, with the residual asymmetric component providing the correct relic density.  If the asymmetry is later washed out (through oscillations), DM annihilation today can give an enhanced $\gamma\gamma$ rate, while $f \bar f$ is $p$-wave or chirality-suppressed.
\end{itemize}
We illustrated these exceptions using simple models, showing in each case that an enhanced $\gamma\gamma$ rate can be naturally reconciled with the correct DM relic density and $\gamma$ continuum constraints.  Clearly, a broad range of model-building possibilities lies within the general framework of these exceptions, beyond the simple models we considered.

Virtually all the models we discussed here have a common feature: the presence of new charged states with mass near the DM mass.  Such charged states would be prime targets for LHC searches, and could play an important role in modification of Higgs couplings to $\gamma \gamma$.  In addition to the line from $\gamma \gamma$ annihilation, there appears to be another, lower energy line around $111$ GeV, which may be consistent with annihilation to $\gamma Z$.  Depending on the $SU(2)_L$ quantum numbers of the charged states generating the effective DM coupling to photons, this line may also arise from a similar process as the one that generates the 130 GeV line.  We leave an exploration of these points for future work.

\vspace{0.5cm}

\section{Acknowledgements}  We thank Geraldine Servant for insightful discussions.  HBY thanks the Center for Theoretical Underground Physics and Related Areas (CETUP* 2012) in South Dakota for its hospitality and for partial support during the completion of this work.  The work of ST, HBY and KZ is supported by the DoE under contract DE-SC0007859.  The work of HBY and KZ is also supported by NASA Astrophysics Theory Grant NNX11AI17G and by NSF CAREER award PHY 1049896.

\bibliography{line}

\end{document}